\documentclass{PoS}

\usepackage{booktabs}

\newcommand{\tr}{\mathrm{tr}\,}
\newcommand{\Real}{\mathrm{Re}\,}
\newcommand{\ev}[1]{\left\langle #1 \right\rangle}
\newcommand{\eq}[1]{Eq.~(\ref{#1})}
\newcommand{\fig}[1]{Fig.~\ref{#1}}

\newcommand{\sect}[1]{Section~\ref{#1}}
\newcommand{\tmtextbf}[1]{{\bfseries{#1}}}

\newcommand{\tmtexttt}[1]{{\ttfamily{#1}}}

\title{Progress in Gauge-Higgs Unification on the Lattice}

\ShortTitle{Gauge-Higgs Unification}

\author{\speaker{\mbox{Francesco Knechtli\thanks{Speaker. FK thanks CERN for hospitality.}}~, Kyoko Yoneyama}, Peter Dziennik\\
        Department of Physics, Bergische Universit{\"a}t Wuppertal\\
        Gaussstr. 20, D-42119 Wuppertal, Germany\\
        E-mail: \email{knechtli@physik.uni-wuppertal.de},
                \email{yoneyama@physik.uni-wuppertal.de},
                \email{dziennik@uni-wuppertal.de}}

\author{Nikos Irges\\
        Department of Physics, National Technical University of Athens\\
        Zografou Campus, GR-15780 Athens, Greece\\
        E-mail: \email{irges@mail.ntua.gr}}

\abstract{
We study a five-dimensional pure $SU(2)$ gauge theory formulated on the 
orbifold and discretized
on the lattice by means of Monte Carlo simulations. The gauge symmetry 
is explicitly broken to
$U(1)$ at the orbifold boundaries. The action is the Wilson plaquette action 
with a modified weight
for the boundary $U(1)$ plaquettes. We study the phase transition and present 
results for the spectrum and the shape
of the static potential on the boundary. 
The latter is sensitive to the presence of a massive Z-boson, in
good agreement with the directly measured Z-boson mass.
The results may support an alternative view of the lattice orbifold 
(stemming from its mean-field study) as a 5d bosonic superconductor.
}

\FullConference{31st International Symposium on Lattice Field Theory - LATTICE 2013\\
		July 29 - August 3, 2013\\
		Mainz, Germany}

\begin{document}

\section{The $SU(2)$ lattice orbifold and its symmetries \label{sec:orbsymm}}

\begin{figure}\centering
  \resizebox{9cm}{!}{\includegraphics[angle=0]{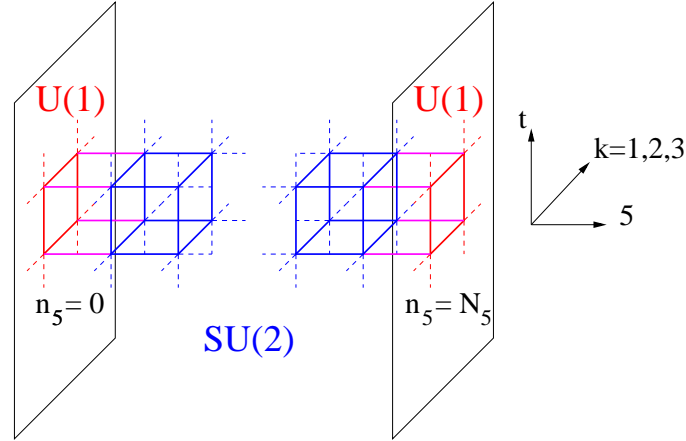}}
  \caption{Schematic representation of the 5d lattice where the $SU(2)$ orbifolded gauge theory is defined.}
  \label{fig:latorb}
\end{figure}

We use a five-dimensional (5d) anisotropic
Euclidean lattice with $T\times L^3\times(N_5+1)$ points.
The lattice spacing is $a_4$ in the four-dimensional (4d) hyperplanes
orthogonal to the extra dimension and $a_5$ along the extra dimension. 
The gauge links $U_M(n)\in SU(2)$ connect the lattice points 
$n+\hat{M}$ with $n$. The Euclidean index $M$ runs over the temporal
$M=0$ and spatial $M=k,5$ ($k=1,2,3$) directions. 
We will also use Greek indices to denote the $\mu=0,k$ directions.
The lattice is assumed to be periodic except for the extra dimension,
which is an interval with boundaries originating from an orbifold projection
\cite{Irges:2004gy}.
The physical length of the extra dimension is $\pi R=N_5 a_5$.
We employ the anisotropic Wilson plaquette action which is defined as
\begin{equation}\label{eq:orb_action}
  S^{\rm orb.}_W =
  \frac{\beta}{2}  \left[ \frac{1}{\gamma}\sum_{{\rm 4d}\;{\rm p}} w\,
  \Real\tr\{I-P_{\mu \nu}(n)\} + 
  \gamma \sum_{{\rm 5d}\;{\rm p}} \Real\tr\{I-P_{\mu 5}(n)\} \right] \,,
\end{equation}
where the weight factor is $w=1/2$ for the boundary plaquettes and
$w=1$ otherwise. Only counterclockwise oriented plaquettes $P_{MN}$
are summed over and $I$ is the identity matrix. The anisotropy parameter
$\gamma$ is in the classical continuum limit $\gamma=a_4/a_5$.
Instead of $(\beta,\gamma)$ we will use the equivalent parameter pair
$(\beta_4=\beta/\gamma,\beta_5=\beta\,\gamma)$.
Along the extra dimension with coordinate $n_5\in[0,N_5]$
the orbifold projection specifies Dirichlet boundary conditions
\begin{equation}\label{eq:orb_bc}
U_\mu(n)\,=\,g\,U_\mu(n)\,g^{-1} \,\Rightarrow\,
U_\mu(n)\,=\,{\rm e}^{\phi(n)\,g}\,\in\,U(1)
\end{equation}
at $n_5=0$ and $n_5=N_5$ with the $SU(2)$ projection matrix
\begin{equation}\label{eq:orb_g}
g=-i\sigma^3 \,.
\end{equation}
Thus, at the interval ends, the gauge symmetry is explicitly broken
to a $U(1)$ subgroup of $SU(2)$ which is left invariant by group
conjugation with $g$.
The lattice of the $SU(2)$ orbifolded gauge theory is schematically 
represented in \fig{fig:latorb}.
There are three types of gauge links: 4d $U(1)$ links contained in the two 
boundaries at $n_5=0$ and $n_5=N_5$, hybrid extra dimensional links (with
one end on a boundary transforming under $U(1)$ and the other end in
the bulk transforming under $SU(2)$) and the remaining bulk $SU(2)$ links.

In addition to the local gauge symmetry, the orbifold theory possesses
the following global symmetries
\begin{equation}\label{eq:global_sym}
Z\otimes F\otimes {\cal F} \,.
\end{equation}
$Z$ is a center transformation in the 4d hyperplanes by an element of the
center $\mathbb{Z}_2=\{\pm I\}$ of $SU(2)$.
$F$ is a reflection with respect to the middle of the orbifold interval
$n_5=N_5/2$.
The fixed point symmetry is ${\cal F}={\cal F}_L\oplus {\cal F}_R$.
On the $L$ (``left'') boundary at $n_5=0$, ${\cal F}_L$ is defined as
\begin{eqnarray}\label{eq:fp_sym_L}
U_5(n_5=0) & \to & g_F^{-1}\,U_5(n_5=0) \\
U_\nu(n_5=0) & \to & g_F^{-1}\,U_\nu(n_5=0)\,g_F
\end{eqnarray}
and on the $R$ (``right'') boundary at $n_5=N_5$, ${\cal F}_R$ is defined as
\begin{eqnarray}\label{eq:fp_sym_R}
U_5(n_5=N_5-1) & \to & U_5(n_5=N_5-1)\,g_F \\
U_\nu(n_5=N_5) & \to & g_F^{-1}\,U_\nu(n_5=N_5)\,g_F \,.
\end{eqnarray}
In order that ${\cal F}_{L(R)}$ are consistent symmetry transformations, 
a boundary $U(1)$ gauge link should remain 
in the $U(1)$ group after conjugation by $g_F$ .
Moreover, the transformations have to commute
with the orbifold projection and $g_F$ has to satisfy \cite{Ishiyama:2009bk}
\begin{equation}\label{eq:orb_consistency}
g\,g_F = g_F\,g\,z_G \,,
\end{equation}
where $z_G$ is an element of the center of $SU(2)$, i.e. $z_G=\pm I$.
If $z_G=-I$ or equivalently
\begin{equation}\label{eq:stick}
\{g,g_F\}=0 \,,
\end{equation}
for example $g_F={\rm e}^{i\theta}(-i\sigma^2)$, the fixed point
transformations ${\cal F}_{L(R)}$ are the stick symmetries introduced in
\cite{Ishiyama:2009bk} and we denote them by
${\cal S}_{L(R)}$. If $z_G=I$ or equivalently $[g,g_F]=0$, then 
the transformations ${\cal F}_{L(R)}$ are global gauge transfomations.

In order to build lattice operators for the scalar (Higgs) and vector 
(gauge boson) particles, see \cite{Irges:2006hg},
we will use the boundary-to-boundary-line
\begin{equation}\label{eq:btobline}
l = \prod_{n_5=0}^{N_5-1}U_5(n_5) \,.
\end{equation}
and the orbifolded Polyakov loop\footnote{The subscript $L$ ($R$)
indicates that an operator is defined on the $L$ ($R$) boundary.}
\begin{equation}\label{eq:orbpoly}
P_L = l g l^\dagger g^{-1} \,.
\end{equation}
Higgs operators, with spin $J=0$, 
charge conjugation $C=1$ and spatial parity $P=1$, are defined by
\begin{equation}\label{eq:higgsop}
\tr P_L \,,\quad \tr \Phi_L\Phi_L^\dagger \,,
\end{equation}
where $\Phi_L = 1/(4N_5)\,[P_L-P_L^\dagger,g]$.
Inspired by \cite{Montvay:1984wy}, we define a gauge boson operator,
with $J=1$, $C=-1$ and $P=-1$, by
\begin{equation}\label{eq:gaugebosonop}
\tr Z_{Lk} \,,\quad Z_{Lk}(n) = g\,U_k(n)\,\alpha_L(n+\hat{k})\,U_k^\dagger(n),\alpha_L(n) \,,
\end{equation}
where $\alpha_L = \Phi_L/\sqrt{\det\,\Phi_L}$.
Note that while $\tr P_L$ is odd under both stick symmetries
${\cal S}_L$ and ${\cal S}_R$,
$\tr Z_L$ is odd under ${\cal S}_L$ and even under ${\cal S}_R$.

Another quantity which will be used in particular to study dimensional
reduction from five to four dimensions is the static potential $V(r)$
extracted from Wilson loops. We take the latter to be defined in a 4d 
hyperplane, so the static potential depends on the coordinate $n_5$.

In order to build a
larger variational basis, the links used to construct the Higgs
and gauge boson operators and the Wilson loops
are smeared with a HYP smearing \cite{Hasenfratz:2001hp} which 
does not use the temporal links and
is adapted for the orbifold. 
In particular the smearing parameters are set to
$\alpha_1=0.5$, $\alpha_2=0.4$ and $\alpha_3=0.2$ and 
the spatial links in the 4d hyperplanes are not smeared
along the extra dimension.

\section{Phase diagram}

\subsection{Mean-field \label{sec:mf}}
\begin{figure}\centering
  \resizebox{10cm}{!}{\includegraphics[angle=0]{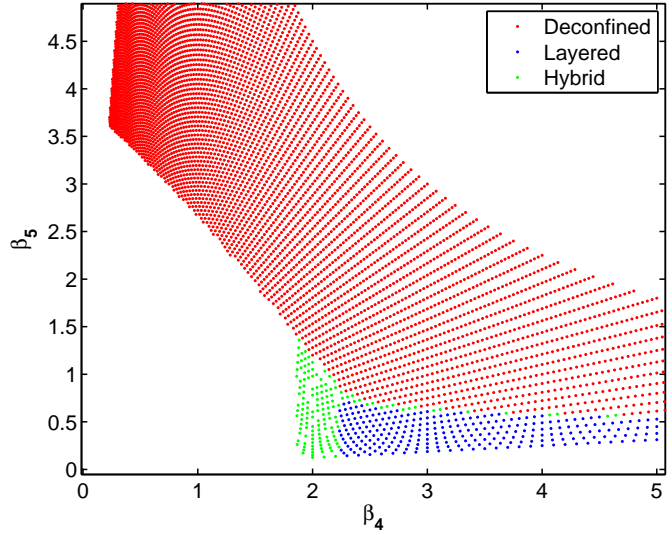}}
  \caption{\small The mean-field phase diagram. The size of the extra dimension
is $N_5=12$. Each point represents the solution of a numerical iterative 
process.}
  \label{fig:mf_phasediagram}
\end{figure}
\fig{fig:mf_phasediagram} shows the phase diagram based on the
solutions obtained for the mean-field background $\overline{v}_0$.
Due to the orbifold boundaries, we distinguish a background
$\overline{v}_0(n_5)$, $n_5=0,1,\ldots,N_5$ in the 4d hyperplanes
and $\overline{v}_0(n_5+1/2)$, $n_5=0,1,\ldots,N_5-1$ along the
extra dimension. For the details of the mean-field formulation we 
refer to \cite{Irges:2012ih}. The
mean-field phase diagram in \fig{fig:mf_phasediagram} has
four phases:
\begin{enumerate}
\item
$\overline{v}_0(n_5)=0$, $\overline{v}_0(n_5+1/2)=0$:
        confined phase (white color);
\item
$\overline{v}_0(n_5)\neq0$, $\overline{v}_0(n_5+1/2)\neq0$:
deconfined phase (red color);
\item  
$\overline{v}_0(n_5)\neq0$, $\overline{v}_0(n_5+1/2)=0$:
layered phase, cf. \cite{Fu:1983ei} (blue color);
\item
$\overline{v}_0(n_5=0)\neq0$, $\overline{v}_0(n_5=N_5/2)=0$:
hybrid phase (green color).
\end{enumerate}
The mean-field sees only bulk phase transitions and is not
sensitive to compactification effects. Approaching the phase
boundary from the deconfined phase, we define the
critical exponent $\nu$ of the inverse correlation length, which is
given by the Higgs mass $M_H=a_4m_H(\beta) = (1-\beta_c/\beta)^\nu$
(and is measured from the Euclidean time correlator of the operator
in \eq{eq:higgsop}). 
For $\gamma>0.6$, we measure $\nu=1/4$,
the Higgs mass $M_H$ and the background do not vanish as the phase
transition is approached. 
This means that the phase transition
is of first order (the lattice spacing does not go to zero).
For $\gamma\lesssim0.6$, the layered phase appears at the phase boundary
and $\nu$ becomes $1/2$.
Moreover the Higgs mass and the background tend to zero approaching the 
boundary of the deconfined/layered phases. This is consistent with a
second order phase transition. Indeed in \cite{Irges:2012mp},
lines of constant physics (LCPs) have been constructed along which
the continuum limit was taken.
In particular on a LCP with $\rho_{HZ} = m_H/m_Z = 1.38$ (which is the current
experimental value) we find that the Higgs mass is finite without
supersymmetry and predict a $Z'$ state of mass $m_{Z'}\simeq1\,{\rm TeV}$.
The mean-field calculations show that a LCP with
$\rho_{HZ} = 1.38$ at $\gamma\ge1$ does not exist.

\subsection{Monte Carlo \label{sec:MC_phasediagram}}
\begin{figure}\centering
  \resizebox{10cm}{!}{\includegraphics[angle=0]{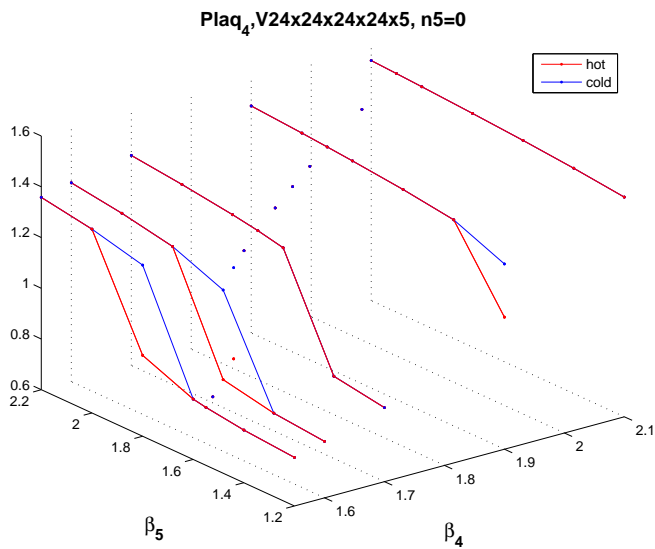}}
  \caption{\small Overview of the Monte Carlo phase diagram of the 4d boundary 
plaquette on the $24^4\times5$ orbifold. Hot (red) and cold (blue) starts.}
  \label{fig:plaq4_phasediagram}
\end{figure}
\fig{fig:plaq4_phasediagram} shows the phase diagram of the 4d plaquette
on the boundary $n_5=0$
in the $(\beta_4,\beta_5)$ plane. We simulated a $24^4\times5$ orbifold
in the parameter region $\beta_4 \in [1.55,2.10]$ and $\beta_5 \in [1.2,2.2]$.
Along the $\gamma=1$ (or $\beta_4=\beta_5$) line we did a finer scan. 
For each parameter point we did two runs (hot and cold start) with 
500 thermalization and 4000 measurement steps. Two steps are separated by 
two iterations of one heatbath and 12 overrelaxation update sweeps. 

There is a bulk phase transition of first order\footnote{
The corresponding transition with periodic boundary conditions along the 
extra dimension (torus) was studied in \cite{Knechtli:2011gq}, see also 
\cite{DelDebbio:2013rka}.}. It is signalled by
a hysteresis. At $\gamma=1$ the hysteresis is between
$\beta=1.60$ and $\beta=1.63$, at slightly smaller $\beta$ values 
than on the torus, cf. \cite{Knechtli:2011gq}.
\begin{figure}\centering
  \resizebox{7cm}{!}{\includegraphics{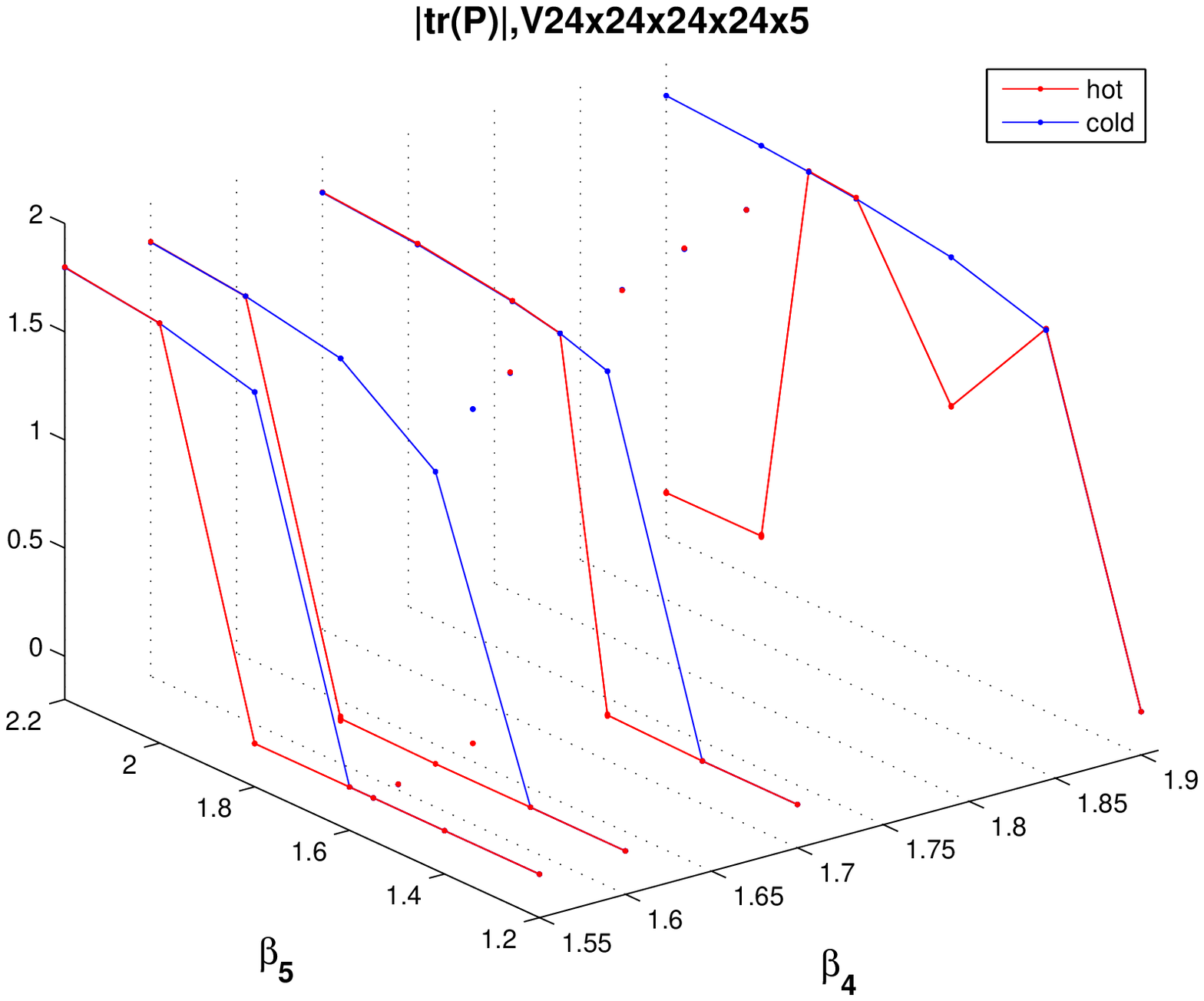}} \ \
  \resizebox{7cm}{!}{\includegraphics{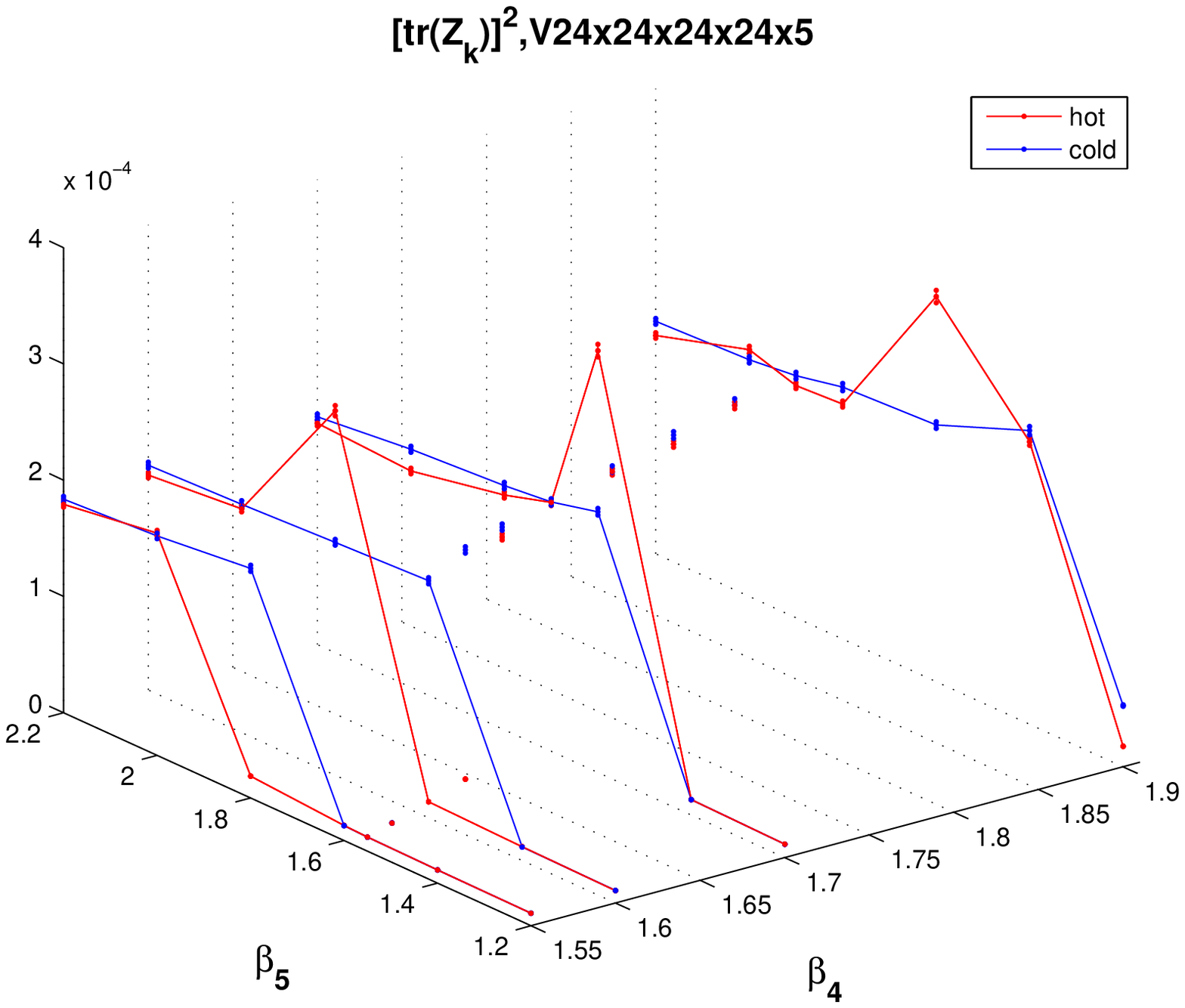}}
  \caption{\small  Overview of the Monte Carlo phase diagram of 
$|\tr P_L|$ (left) and $1/3\sum_k(\tr Z_{Lk})^2$ (right) on the $24^4\times5$ 
orbifold. Hot (red) and cold (blue) starts.}
  \label{fig:poly5_phasediagram}
\end{figure}

\fig{fig:poly5_phasediagram} shows the phase diagram of the absolute value of 
$\tr P_L$ defined in \eq{eq:orbpoly} (left plot) and of 
$1/3\sum_k(\tr Z_{Lk})^2$ defined in \eq{eq:gaugebosonop} (right plot).
The lattices are $24^4\times5$, the same as in \fig{fig:plaq4_phasediagram},
and the links entering the operators are smeared by 10 iterations 
of HYP smearing.
Vertical lines are plotted to show the statistical errors.
Comparing to \fig{fig:plaq4_phasediagram}, we see that both operators
show an hysteresis at the first order bulk phase transition.
The operators $\tr P_L$ and $(\tr Z_{Lk})^2$ have finite volume effects
when $\beta_4$ becomes large
(which we know from a comparison with $16^4\times5$ lattices)
and we therefore restrict the range of parameters plotted.
The irregular behavior at $\beta_4=1.9$ is probably due to
finite volume effects.

A more detailed investigation of the phase diagram is under way.

\section{Spectrum}

On an isotropic ($\gamma=1$) orbifolded $64\times32^3\times5$ lattice
we measure 
the spectrum close to the bulk phase transition ($\beta=1.66,\,1.68$) using
the operators defined in \eq{eq:higgsop} and \eq{eq:gaugebosonop}.
The statistics is of 2000 measurements separated by two update steps (each
update step consists of one heatbath and 16 overrelaxation sweeps).
The variational basis is constructed using smeared links with 5, 15 and 30
HYP smearing iterations.

\begin{figure}\centering
  \resizebox{7cm}{!}{\includegraphics{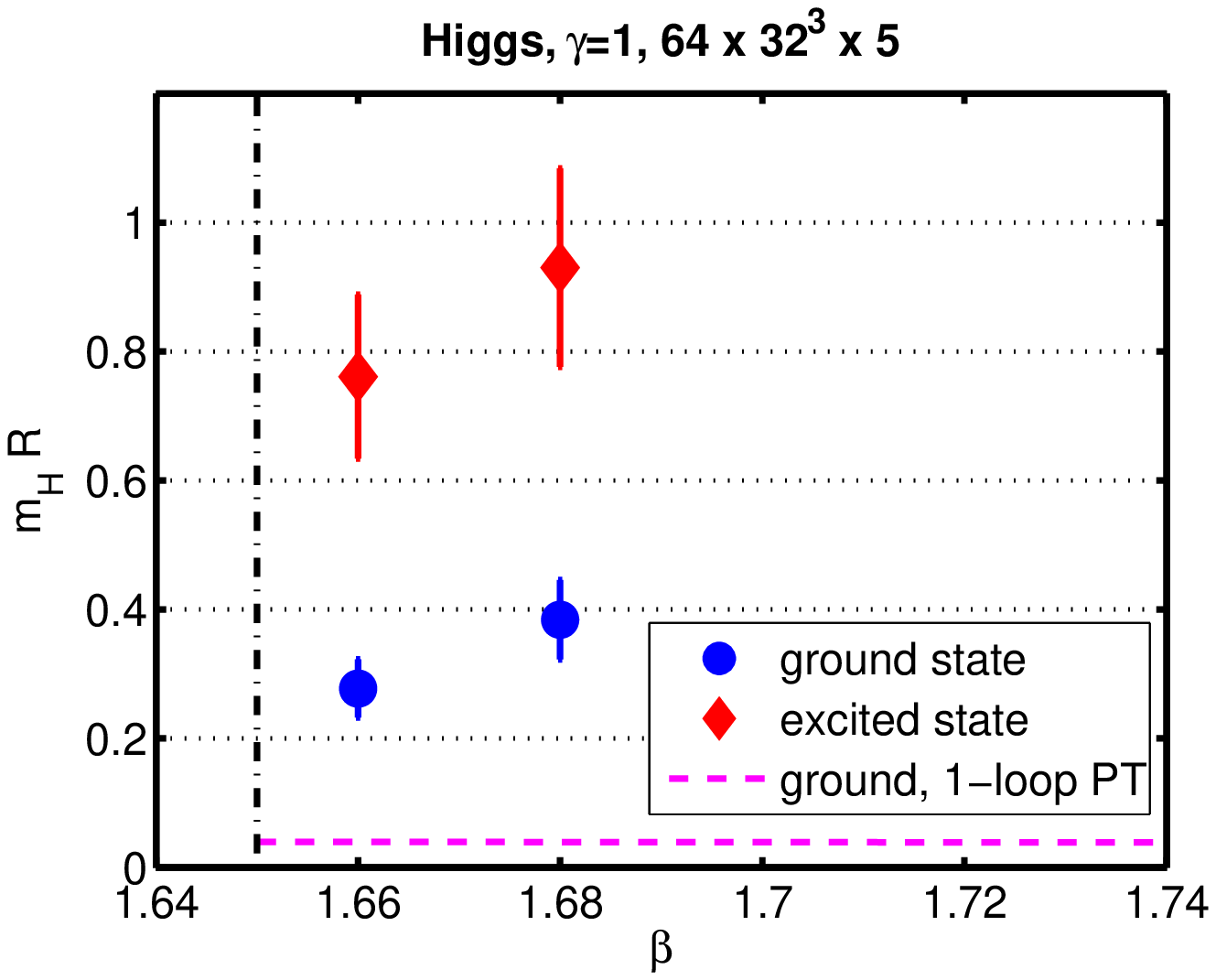}} \ \
  \resizebox{7cm}{!}{\includegraphics{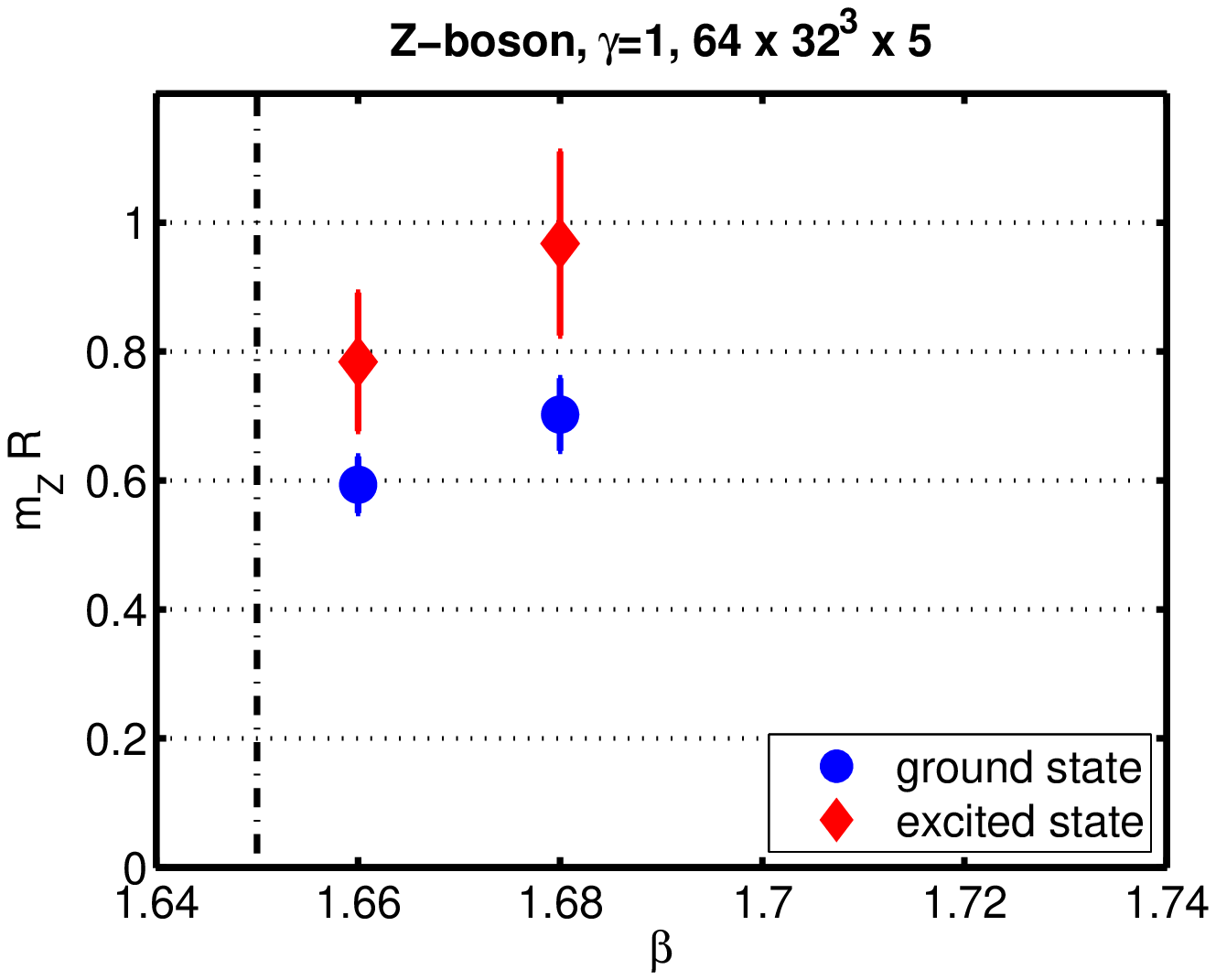}}
  \caption{\small The spectrum for $\gamma=1$, $L=32$ and $N_5=4$ close to
the bulk phase transition marked by the vertical dash-dotted line.
The masses of the Higgs (gauge boson) ground and excited states are shown 
on the left (right) plot in units of $1/R$. 
The Higgs mass is much higher than its value from the
one-loop (continuum) formula \cite{vonGersdorff:2002as,Cheng:2002iz}.}
  \label{fig:masses_g1}
\end{figure}

The masses of the Higgs and gauge boson in units of $1/R$
are shown in \fig{fig:masses_g1}.
Together with the ground states, also the first excited states were resolved.
We find a nonzero gauge-boson mass $m_Z\neq0$
and the value of $m_Z$ does not decrease with the lattice size $L$.
This implies that the Higgs mechanism is at work.
The mass hierarchy is not the one measured by the experiments at CERN, 
we find $m_Z > m_H$.
The masses of the excited states of the Higgs and the Z-boson are 
approximately equal.

\begin{figure}\centering
  \resizebox{7cm}{!}{\includegraphics{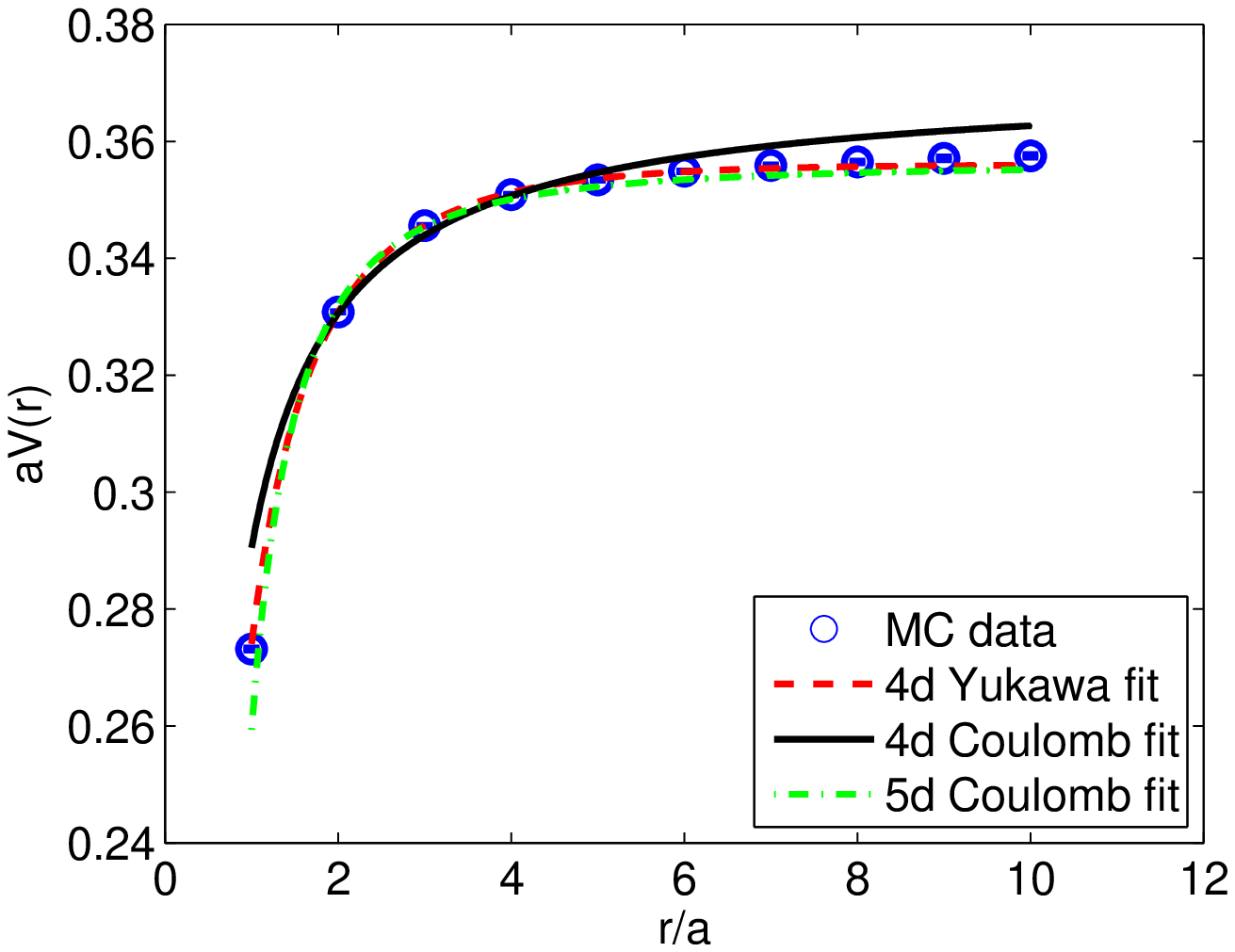}} \ \
  \resizebox{7cm}{!}{\includegraphics{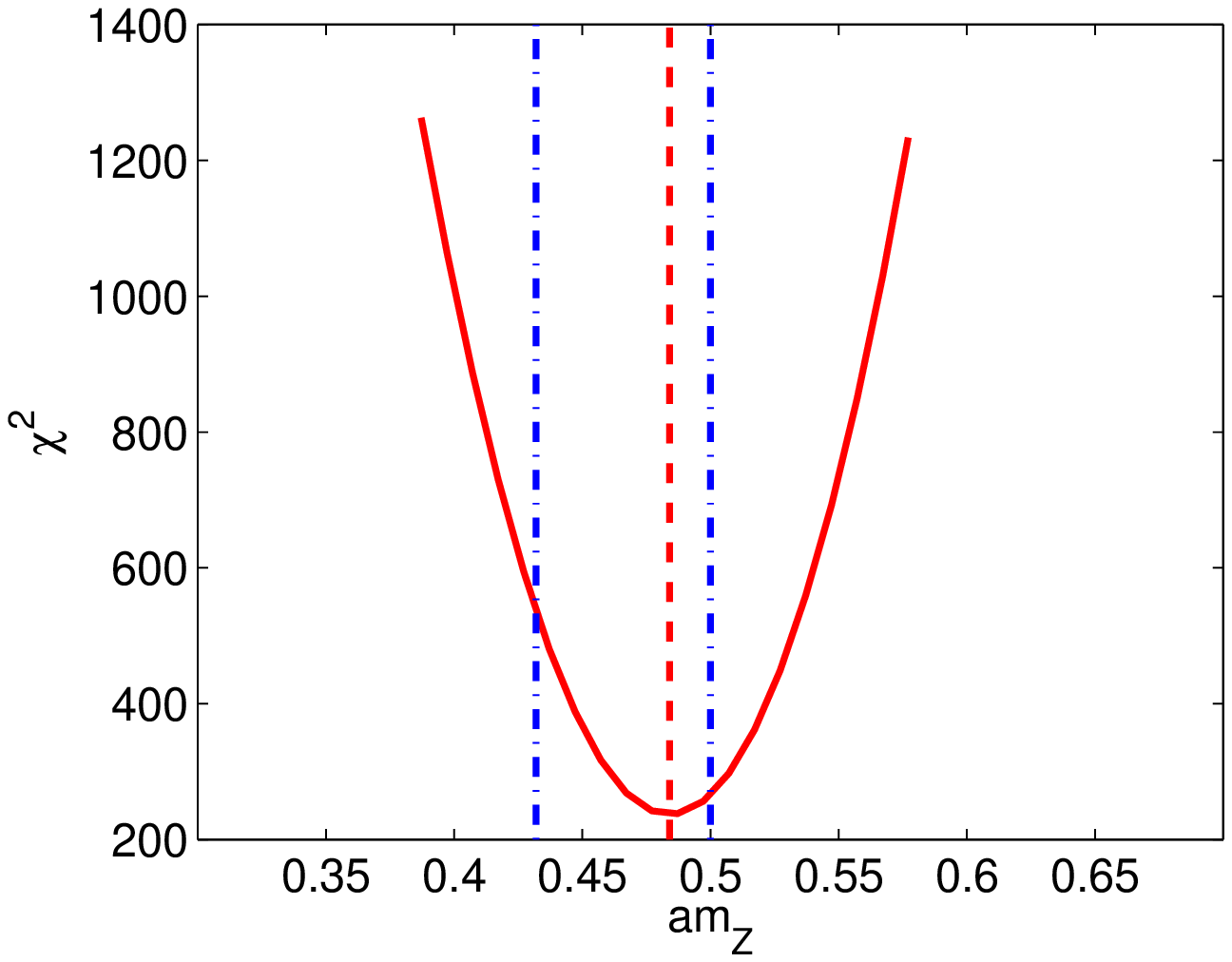}}
  \caption{\small The boundary static potential at $\gamma=1$, $L=32$, $N_5=4$
and $\beta=1.66$ (left plot) favors the 4d Yukawa global fit. The Yukawa mass
which minimizes the $\chi^2$ (right plot, red vertical dashed line) agrees 
well with the directly measured value marked by the
vertical blue dashed-dotted lines.}
  \label{fig:potential_g1_b1p66}
\end{figure}

At $\beta=1.66$ we also measured the static potential on the boundary.
The data (blue points) are shown on the left plot of
\fig{fig:potential_g1_b1p66}.
In order to get a qualitative understanding of the shape of the potential,
we perform 4d Yukawa (red dashed line), 4d Coulomb (black continued line)
and 5d Coulomb (green dash-dotted line) global fits\footnote{
The best 5d Yukawa fit 
has a zero Yukawa mass and is therefore not considered.}. 
In the Yukawa non-linear fit, the Yukawa mass is a fit parameter. 
The preferred global fit is the 4d Yukawa.
The $\chi^2$ of the 4d Yukawa fit is shown on the right plot of
\fig{fig:potential_g1_b1p66} as function of the Yukawa mass $am_Z$.
The value of $am_Z$ which minimizes the $\chi^2$ is consistent with
the directly measured value of $am_Z$, cf. \fig{fig:masses_g1}.

After we proved the existence of the Higgs mechanism, by
finding a nonzero mass for the $U(1)$ gauge boson,
the question of its origin arises.
In particular Elitzur's theorem \cite{Elitzur:1975im} tells that
only global symmetries can be spontaneously broken on the lattice and
phase transitions are characterized by gauge invariant order parameters.
In \sect{sec:orbsymm} we have identified global symmetries, the fixed
point symmetries ${\cal F}_L$ and ${\cal F}_R$, which contain 
the stick symmetries ${\cal S}_L$ and ${\cal S}_R$.
In the deconfined phase, where we measure the masses, the stick
symmetries are spontaneously broken, see \sect{sec:MC_phasediagram}.
This breaking induces the breaking of the other global symmetries,
which are global gauge transformations, this is the origin of the
Higgs mechanism.
The Polyakov loop $\tr P_L$ in \eq{eq:higgsop} is the order parameter
for confinement/deconfinement.
The deconfinement phase can be a Coulomb or a Higgs phase.
We conjecture that the operator $\tr Z_{Lk}$ in \eq{eq:gaugebosonop},
is the order parameter of the Higgs phase, namely the Higgs mechanism
happens when $\ev{\tr Z_{Lk}^2}\neq0$.
\begin{figure}\centering
  \resizebox{6cm}{!}{\includegraphics[angle=0]{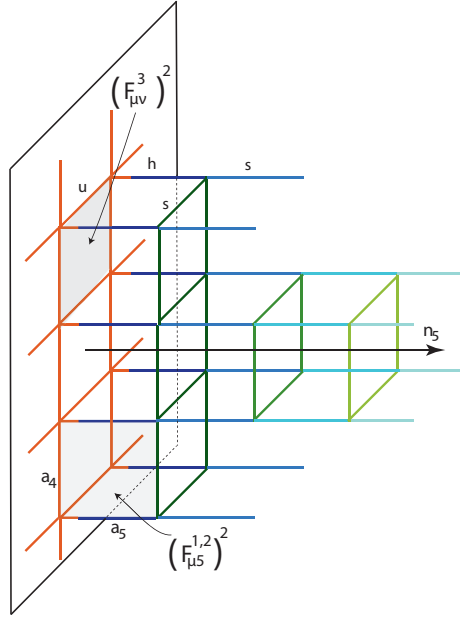}}
  \caption{The mean-field background defines a ``crystal'' in coordinate
and gauge space. Links of different colors correspond to different values
of the background.}
  \label{fig:background}
\end{figure}

The Higgs mechanism on the orbifold seems to have a different origin
than the Hosotani mechanism \cite{Hosotani:1983vn}, which was formulated
in perturbation theory and works only when fermions are included, see also
the lattice study of \cite{Cossu:2013ora}.
The orbifold mechanism of spontaneous symmetry breaking
could be related to a bosonic superconductor.
The mean-field background breaks translation invariance,
like a crystal, see \fig{fig:background}. 
Gauge fluctuations around the mean-field background are 
like phonons, the Polyakov loop (Higgs) with $U(1)$ charge $2$ is like a 
Cooper pair and the Higgs mechanism, due to gauge-Higgs interaction, happens 
like in a superconductor slab, where the photon becomes massive.

\section{Conclusions}

We have presented a Monte Carlo study of the 5d $SU(2)$ orbifold,
extending previous results from \cite{Irges:2006hg}.
A second order phase transition is not found. Instead
there is a line of bulk first order phase transitions and a new transition
for anisotropy $\gamma<1$ which is
signalled by the boundary (but not by the bulk) 4d plaquette.
The boundary transition is also of first order where we did simulations.

The spectrum is measured so far only at $\gamma=1$ for an orbifold with
$N_5=4$, where we find a massive gauge boson which is heavier than the 
Higgs scalar.
The mass of the gauge boson is consistent with the mass extracted from
a 4d Yukawa fit to the boundary static potential, thus supporting
dimensional reduction.
It remains to be seen how the mass hierarchy is at $\gamma<1$, where
in the mean-field calculation it was possible to reproduce the
experimentally measured masses for the Higgs and gauge boson.

\end{document}